# Empirical study of software quality evolution in open source projects using agile practices

Alessandro Murgia[1], Giulio Concas[1], Sandro Pinna[1], Roberto Tonelli[1], Ivana Turnu[1],

[1]*Dept. Of Electrical and Electronical Engineering, University of Cagliari, piazza d'Armi, 09123 Cagliari, Italy*
{alessandro.murgia, concas, sandro.pinna, roberto.tonelli, ivana.turnu,}@diee.unica.it

***SUMMARY.***

**We analyse the time evolution of two open source Java projects: Eclipse and Netbeans, both developed following agile practices, though to a different extent. Our study is centered on quality analysis of the systems, measured as defects absence, and its relation with software metrics evolution. The two projects are described through a software graph in which nodes are represented by Java files and edges describe the existing relation between nodes. We propose a metrics suite for Java files based on Chidamber and Kemerer suite, and use it to study software evolution and its relationship with bug count.**

## 1. INTRODUCTION

The classic software development cycle is composed of a series of sequential activities such as requirements analysis, architecture design, low-level design, coding, testing, delivery, and eventually maintenance. An agile process development, on the other hand, is characterized for being iterative with short iterations, with almost all quoted activities performed, to some extent, in each iteration throughout the project.

The agile approach in software production is quite recent, but the number of projects which already use it is a hint of its goodness. However, quantitative assessments of the effectiveness of agile methodologies (AMs) is still subject of research. Note that real projects usually apply a mix of agile practices rather than a specific, well defined AM, "following the book". AMs contribution to software development could be highlighted in the quality improvement that it can obtain. In fact goals of AMs are to satisfy the customer through continuous delivery of valuable software and also to welcome changing requirements (http://agilemanifesto.org).

Unfortunately, quality is not easily measurable. However, the number of defects can be used as a key indicator of product quality. So we deem that a suitable assessment of the effectiveness of agile practices can be the study of defect reduction during product development.

Defect distribution and software evolution over time are still research topics. Their understanding is needed to effectively organize software production. Andersson and Runeson (2007) discussed the Pareto distribution of bugs in modules, without entering into the details of the statistical properties of software which determine such property. Recently, a work of Zhang (2008), based on Anderson's studies, analyzes again the bug distribution across compilation packages in Eclipse, finding a Weibull distribution. Some aspects of software evolution and its relation with bug introduction are studied by Kim *et al.* (2006) and by Purushothaman and Perry (2005). Kim inspected "micro-pattern" evolution in Java classes identifying which of them is more bug-prone. Purushothaman analyzed the software development process to identify what are the relationships between small changes to the code and bug growth.

Finally, Zimmermann and Nagappan (2008) performed a network analysis on dependences graphs, built on binary files, and how dependencies correlate with, and predict, defects.

These works pay attention on defect introduction in software system, but study of defects inside a graph and an analysis of graph evolution is, to our knowledge, completely missing.

This work presents a study of software structure and bug dynamics matching recent advances on object-oriented source code analysis and representation using complex network theory reported in (Concas *et al.* 2007), with an analysis of bug fixing activities as reported in configuration management systems – such as Concurrent Versions System (CVS) – and in bug reporting repositories – such as Bugzilla. The study is performed on two large open source projects, Eclipse (2009) and Netbeans (2009). Eclipse is well known to be developed using agile practices such as test-driven development and refactoring since its beginning. Netbeans team used only partially automatic testing, though at a rate increasing with subsequent versions, and does not practice systematic refactoring. One of our goals is to evaluate if the different usage level of such practices has some impact on bug distribution and dynamics.

Since bug management operations are made and recorded on code files, and not on single classes, we introduce new code metrics, based on Chidamber and Kemerer's (CK) suite reported in (Chidamber and Kemerer 1998), but referring to object-oriented code files. These metrics are matched with bug information as found in CVS and Bugzilla logs, properly elaborated to filter out "false positives".

The results of the comparison between Eclipse and Netbeans are presented and discussed, showing that Eclipse has a higher regularity in project flow and bug behavior, though definitive conclusions cannot be drawn.

The remainder of the paper is organized as follows. In section 2 we describe how to build the software graph related to system software, how to extend the definition of CK metrics to compilation units (CU), and finally how to use issue-tracking systems to locate bug associated to CUs. In section 3 we present the two Java systems studied and their main features. In section 4 we present and discuss the results of the analysis. In section 5 we argue about threats to validity and future works. The paper ends with conclusions in section 6.

## 2. METHOD

Our approach focuses on a static code analysis of object-oriented code, Java in this case. The key concepts of Object Oriented (OO) code are classes and interfaces. The most important metrics which deal with source code classes are CK suite, and the MOOD suite reported in (Abreu 1995). More recently, some authors (Concas *et al.* 2007) introduced the concept of OO class graph, an oriented graph whose nodes are the classes – and possibly the interfaces – of the system, and whose edges are the relationships between classes, namely inheritance, composition and dependence.

The number and orientation of edges allow to study the coupling between nodes. In this graph the in-degree of a class is the number of edges directed toward the class, and measures how much this class is used by other classes of the system. The out-degree of a class is the number of edges leaving the class, and represents the level of usage the class makes of other classes in the system. We also associate to each node of the graph the values of the OO metrics computed on the class represented by it, and more specifically the four most relevant CK metrics:

1. Weighted Methods per Class (WMC). A weighted sum of all the methods defined in a class. We set the weighting factor to one to simplify our analysis.

2. Coupling Between Objects (CBO). The counting of the number of classes which a given class is coupled to.

3. Response For a Class (RFC). The sum of the number of methods defined in the class, and the cardinality of the set of methods called by them and belonging to external classes.

4. Lack of Cohesion of Methods (LCOM). The difference between the number of non cohesive method pairs and the number of cohesive pairs.

Furthermore, we computed the lines of code of the class (LOC), excluding blanks and comment lines. This is useful to keep track of CU dimension because it is known that a "long" class is more difficult to read than a short class.

When dealing with bugs, however, this approach presents the issue that bug-tracking systems, such as Bugzilla, work on source code files and not on single classes. We know that every system class resides inside a CU, which is a Java file. While most files include just one class, there are files including more than one class. For instance, in Eclipse 10% of CUs host more than one class, whereas in Netbeans this percentage is 30%. Bugs and bug fixing always refer to CUs. To make consistent bug tracking with source code, we decided to extend CK metrics from classes to CUs.

CUs represent therefore the main element of our study. So, we defined a CU graph whose nodes are the CUs of the system. In this graph, two nodes are connected with a directed edge if at least one class inside the CU associated with the first node has a dependency relationship with one class inside the CU associated with the second node. The methods of all the classes contained in the CU become CU-methods. We refer to this graph for computing in-links and out-links of a node. Taking into account this graph, we reinterpreted our metrics from classes to CUs as follows:

- CU LOCS is the sum of the LOCS of classes contained in the CU.

- CU CBO is the number of out-links of each node, excluding those representing inheritance. This definition is consistent with that of CBO metrics for classes.

- CU LCOM and CU WMC are the sum of LCOM and WMC metrics of the classes contained in the CU, respectively.

- CU RFC is the sum of all distinct methods of the CU, plus all the disctinct and external methods possibly called by the formers.

For each CU we have thus a set of 6 metrics: In-links, Out-links, CU LOCS, CU LCOM, CU WMC, CU RFC and CU CBO.

Once the CU graph is computed, we have to find which nodes are hit by bugs. To obtain this information, it is necessary to check the configuration management system – CVS in our case – log file, and the data contained in the issue-tracking system, such as Bugzilla.

We consider a CU as affected by a bug when it is modified for bug fixing. All fixing activities performed on the system by its developers are recorded on the CVS log. In fact, all commit operations are tracked in the CVS log as single entries. Each entry contains various data, among which the date, the developer who made the changes, an annotation referring to the reasons of the commit, and the list of CUs interested by the commit. In the case the commit is associated to a bug fixing activity, this is written in the annotation, though not in a standardized way.

It is not simple to execute a correct mapping between a bug and the CU(s) hit by it. Some researches where this issue was tackled are reported in (Śliwerski *et al.* 2005) (Fischer *et al.* 2003). In our approach, we first analyzed the CVS log, to locate commit messages associated to fixing activities. Then, the extracted data are matched with information found in the issue-tracking system. Relying only on the analysis of fixing messages is insufficient – often programmers describe ambiguously the correction made, and which bugs were fixed. Each bug is identified by a whole positive number (ID). In commit messages it can appear a string such as "Fixed 141181" or "bug #141181", but sometimes only the ID is written. Clearly , every integer positive number is a potential bug. On the other hand, if we labeled each ID as a bug, we would consider as bug ID numbers bearing a complete different meaning. To cope with this issue, we applied the following strategies:

1. we considered only positive integer numbers present in the issue tracker as valid bug IDs related to the same release;
2. we did not consider some numeric intervals particularly prone to be a false positive bug ID.

The latter condition is not particularly restrictive in our study, because we do not consider the first releases of the studied projects, where bugs with "low" ID appear.

All IDs not filtered out are considered bugs and associated to the addition or modification of one ore more CUs, as reported in commit logs. The total number of bugs hitting a CU in each release constitutes the bug metric we consider in this study.

Some threats to the calculation of this metric are: possible wrong mapping between bug and CU due to typos in the bug ID inserted in the message; or changes of the fixing release (in issue tracking) after the mapping. However, since these events are rare, we don't consider them as invalidating for our study.

Note that a "bug" reported in a bug management system has a broad sense. It may denote a true error in the code, but also an enhancement of the system, or fixing a requirement error, not a coding error. Moreover, when many CUs are affected by a single bug, it is possible that some of them are in fact modified not because they are faulty, but as a side-effect of modifications made in other CUs. So, when we talk of a CU "hit" by a bug, this does not necessarily mean that the CU included and error, or was poorly coded.

## 3. THE CASE STUDIES

Our case studies are two very large Java projects in the domain of integrated development environment - Eclipse and Netbeans. Both are mature and successful software tools derived from open source projects, with years of development. We choose these projects because they give free access to their CVS repository, a key factor for our research, because we need complete access to the source code of their various versions. Another important property of these two projects are theirs similarities, which are useful to compare them.

The development of both analyzed projects proceeds by main releases and patching releases. Main releases (MR) are denoted by a two-digit decimal number – such as 3.2 o 4.0 – and represent a major step in functionalities. Patching releases (PR) are denoted by a third number following the name of the MR they are updating – for instance 3.2.1 is a PR of MR 3.2. PR do not add substantial features, but fix the bugs and other issues found in previous versions of the release. After a MR is released, its source code is used for fixing bugs through "patches", resulting in subsequent PRs. The work to produce a new MR, on the other hand, proceeds in parallel, and when another MR is released, its source code may be quite different from the code of the previous MR. Note that some MR do not have PR, but directly evolve into another MR. In both examined projects, there is approximately a MR every 8-12 months. It is possible to see that Eclipse development process is more regular than Netbeans, because almost each MR is delivered after a year and each MR has two or three associated PRs. In Netbeans MRs follow a more changeable delivery.

In our study, we considered all the source code available on CVS repositories, that is the core system and the main add-ons. Table 1 shows the number of CUs of the considered MRs of both Eclipse and Netbeans projects.

**Table 1.** The releases of Eclipse and Nebeans Java IDEs, with their CUs.

| Eclipse | | | Netbeans | | |
|---|---|---|---|---|---|
| Main Release | Nr. of CUs | Release date | Main Release | Nr. of CUs | Release date |
| 2.1 | 7885 | 03-2003 | 3.1 | 2420 | - |
| 3.0 | 10584 | 06-2004 | 3.2 | 3350 | - |
| 3.1 | 12174 | 06-2005 | 3.3 | 4421 | - |
| 3.2 | 13221 | 06-2006 | 3.4 | 6282 | - |
| 3.3 | 14564 | 06-2007 | 3.5 | 7391 | 06-2003 |
| | | | 3.6 | 8350 | 04-2004 |
| | | | 4.0 | 9365 | 12-2004 |
| | | | 4.1 | 11768 | 05-2005 |
| | | | 5.0 | 12137 | 01-2006 |
| | | | 5.5 | 15970 | 10-2006 |
| | | | 6.0 | 37145 | 12-2007 |

Both Eclipse and Netbeans are huge projects from the perspective of the number of CUs. In Table 1 we do not report data about PRs, but they are always almost equal to those of the corresponding MRs. Both system exhibit a steady growth in their number of CU.

Note that Eclipse project has a very regular trend, with each MR followed by at least one PR. We know that in this project several agile practices have been used since its inception, in particular test-driven development and refactoring.

Netbeans, on the other hand, had a less regular development. Many MRs are not followed by any PR. Moreover, the adoption rate of agile practices was lower than in Eclipse. In particular, private communications with some members of the project highlighted that automatic testing started in the development of release 3.1, but full coverage is not yet obtained today. Minor refactoring is practiced from time to time. On the other hand, a Behavior/Feature Driven Development is used to control the development process, as well as continuous integration.

## 4. RESULTS

In this section we study Eclipse and Netbeans together, to compare their software development process.

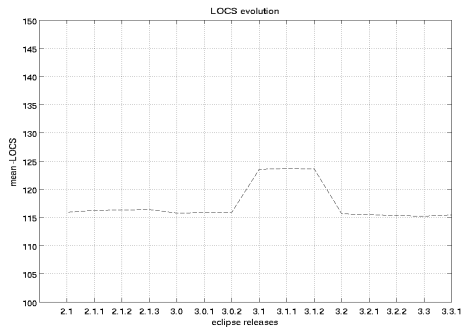

Fig. 1. Mean LOC evolution in Eclipse

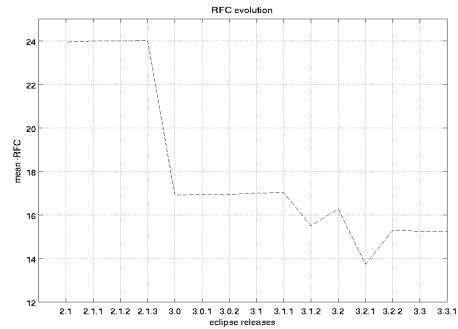

Fig. 2. Mean RFC evolution in Eclipse

Let us first examine the Eclipse project. Figures 1 and 2 illustrate the evolution of the mean value of LOC and RFC, averaged on CUs. We observe that the values of mean LOC are quite stable and then foreseeable, with a peak in release 3.1, that disappears in release 3.2. This means that, with the partial exception of MR 3.1, the addition of new features to the system was systematically achieved adding new CUs and not overloading existing CUs with new functionalities.

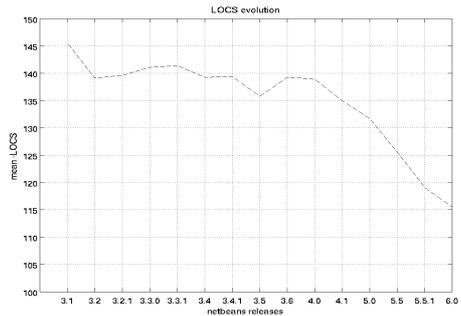

Fig. 3. Mean LOC evolution in Netbeans

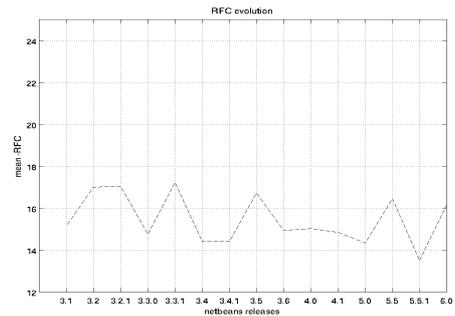

Fig. 4. Mean RFC evolution in Netbeans

The values of RFC show a strong decrease from releases 2 to 3, and then a decreasing trend, though to a lesser extent. We recall that object-oriented metrics are measures of complexity, and thus usually a low value is "better" than a high one.

Figures 3 and 4 show LOC and RFC means in Netbeans. LOC tend to decrease, especially after MR 4.0, showing that the increasing number of CUs (see Table 2) is partially compensated by a smaller average code length of CUs. The mean RFC value is quite variable between 13 and 17, and does not exhibit a decreasing trend, as is Eclipse. We do not consider CBO evolution because it shows a trend similar to RFC and does not provide new information. The others CU metrics will be reported in future development of this study.

We can interpret the results on average LOC imagining that new features introduced in the system, are not inserted into CUs already present, but are hosted inside new CUs. With the approximation that each CU hosts only a class, this mean that the responsibility of a class does not change during the time. Such approximation is not too restrictive. In fact CUs containing more than one class are less than 10% in Eclipse and less than 30% in Netbeans. Thus, Eclipse developers appear to have a better expertise with respect to the correct distribution of classes responsibilities. As Eclipse evolves, RFC tends to reduce, despite CU growth. For the single

CU, the mean number of out-links decreases, although the number of nodes available for linking increases.

With regard to Netbeans, observing Figure 3, we find a different behavior. The mean value of LOC per CU decreases with time, and the number of CUs strongly increases. Our hypothesis, which will be analyzed in future works, is that the introduction of new CUs seems to have a double function: to host new features and to take responsibilities already assigned to old CUs in the past. Another possibility to explain this phenomena is a directive, created by development team after release 4.0, which constrains programmers to limit the dimension of new CUs. This particular phenomena will be studied in future work.

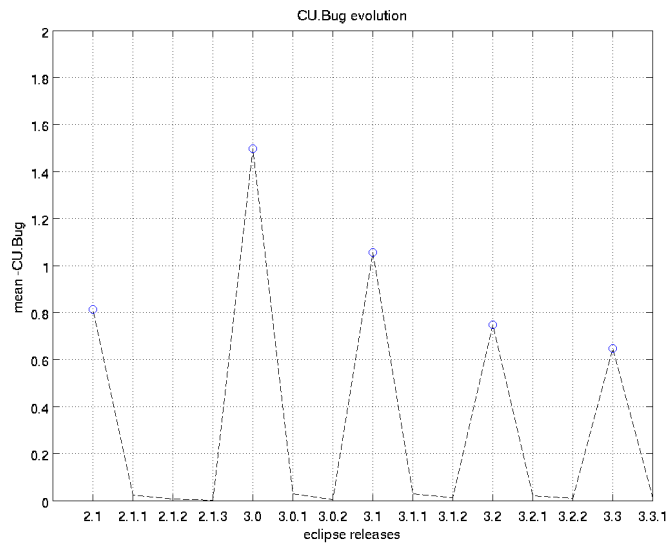

Fig. 5. Evolution of mean number of bugs hitting a CU in Eclipse.

One of the most important aspects of software quality is the absence or reduction. To understand if there is any relationship between the development process adopted by programmers and changes in the system faultiness we analyzed release by release the bug mean value for CUs.

Fig. 5 shows that the mean value of issues for a CU in Eclipse system has a regular trend. For each MR the bug number is high, while in the following PRs the bug number decrease conspicuously. The first PR which follows an MR reduces drastically the bug number. Then this number tends to zero with the second (or third) PR. For this reason a lot of defects are signaled on it. In following PRs the software house, tries to solve them as best as possible. Thus the number of defects discovered decreases quickly. More than one PR is needed because some bugs are discovered also after a PR is delivering. The figure shows also a clear, steady reduction, with time, of the mean number of bugs hitting a CU, starting from release 3.0.

To understand which type of relation exists between LOC (or RFC) metric and faultness, we compare Fig. 5 with the Figures 1 and 2.

We notice that the increment (or decrement) of the mean number of bugs, is not directly related to a positive (or negative) change of a metric. The same conclusions are valid if we consider WMC, CBO or LCOM metrics. We can conclude that a single metric is not able to justify the bug number variation in Eclipse.

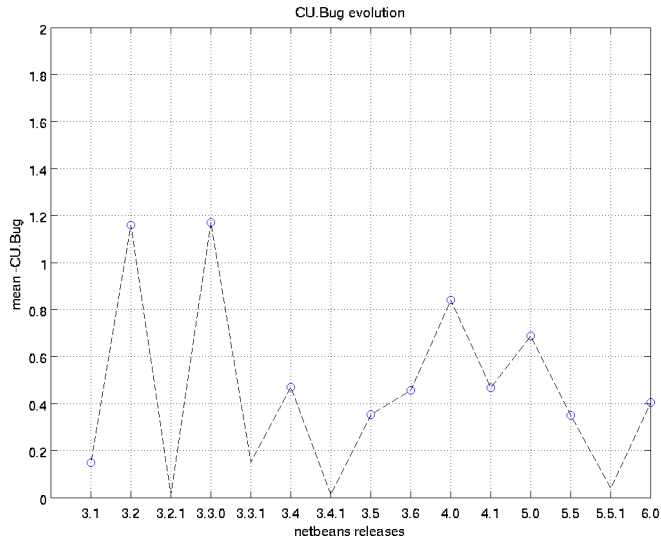

Fig. 6. Evolution of mean number of bugs hitting a CU in Netbeans.

Netbeans exhibits a bug trend similar to Eclipse. In Figure 6, we can observe the pairs of MR and PRs 3.2-3.2.1, 3.3-3.3.1 and 3.4-3.4.1. For each pair, there is first an increment of the number of bugs in MR and then a decrement in PR. This behavior is the same exhibited by MR-PR pairs in Eclipse. In Netbeans, however, there is a strong variability in the bug number during its evolution, and no apparent steady reduction of the mean number of bugs per CU even considering only MRs, as in Eclipse (MRs are indicated by circles in Figure 6). Comparing Fig. 6 with Figures 3 and 4, we notice that also for Netbeans there is no specific metric which is strongly related with the bug number evolution during the time.

To decide where to intensify the efforts in bug search and fixing, it is very important to understand how bugs are distributed. A log-log graph plot of the bug distribution, shows a straight line, meaning a powerlaw behaviour. Thus, there are few CUs hosting the majority of bugs, and most other CUs with a very few bugs, if any. Pareto law is a consequence of this phenomena: 80% of bugs are hosted in less than 20% of CUs.

**Table 2.** Percentage of CUs which contain 80% of bugs

| Eclipse | | Netbeans | |
|---|---|---|---|
| Main Release | Perc. Of CUs | Main Release | Perc. Of CUs |
| 2.1 | 20% | 3.1 | 8% |
| 3.0 | 20% | 3.2 | 20% |
| 3.1 | 19% | 3.3 | 17% |
| 3.2 | 16% | 3.4 | 16% |
| 3.3 | 15% | 3.5 | 14% |
| | | 3.6 | 12% |
| | | 4.0 | 13% |
| | | 4.1 | 8% |
| | | 5.0 | 12% |
| | | 5.5 | 8% |
| | | 6.0 | 8% |

In Table 2 we report, for each MR of both Eclipse and Netbeans, the percentage of CUs containing 80% of bugs. We see that Pareto law clearly holds. We do not report PR statistics, because in PRs the bug number is much smaller and the data would be meaningless. In general, a small subset of the entire CU population contains the large majority of bugs. The implications from a software engineering perspective are clear: if it is possible to identify in advance which CUs subset is at risk, we might improve the system quality at lower cost that searching all CUs for bugs with the same effort.

To summarize the results of our research, we found the following similarities in both Eclipse and Netbeans systems:

- Main Releases may introduce significant changes in the considered metrics. These changes, associated to the introduction of new features, are followed also by the introduction of a significant set of new bugs.

- Patching Releases exhibit a strong stabilization of metric values, and a substantial lower number of bugs. This means that, in both considered systems, the bug fixing activity occurring after a MR has been released, is easy to make.

- There is not a single metric which can explain the bug evolution during the system evolution. OO metrics can be more on less correlated with bugs, but no metric can be used as a strong predictor of them. Metrics, however, are considered indicators of "good" system architecture, and thus might be related to the ease of refactoring and evolving the system. More work is needed to assess this hypothesis, also under a software network perspective.

- In both projects, Pareto law holds. In fact, 20% - 15% or even 8% of CUs contain 80% of all bugs found in the system. This fact, already highlighted by other researchers in other systems, has deep implications from a software engineering perspective.

Eclipse and Netbeans projects bear also some important differences. Remember that a key assumption of this study, supported by communications with some members of both projects, is that Eclipse is a project that has been developed using consistently agile practices, such as Test Driven Development (TDD), refactoring and feature-driven development throughout its life-cycle. Netbeans, on the other hand, applied just feature-driven development throughout its life-cycle, while automatic testing was introduced much more gradually. We believe that at least some of the differences highlighted in our study can be due to this different adoption level of agile practices. Clearly, this evidence is just anecdotal, and more work is needed to quantitatively assess such hypothesis. In short, the main differences between Eclipse and Netbeans are:

- Eclipse metric behavior looks quite stable throughout the project evolution. OO metric average values computed on CUs tend to decrease with time, denoting that refactoring activities work, even in the presence of a system growing both in size and functionalities.

- Average bug count of CUs in MRs steadily decreases, starting from release 3.0, showing that developers are in control of system quality. This effect might be due to the systematic use of TDD, that endows the system a growing asset in term of automatic tests, as its size increases. This tends to catch bugs during development, thus producing main releases with less bugs to find and fix.

- The Eclipse development process is characterized by a higher regularity in terms of MRs and PRs with respect to Netbeans.
- Netbeans project, while characterized by an average lower number of bugs per CU with respect to Eclipse, do not exhibit a foreseeable behavior in RFC metric, and also in the number of bugs per CU, as the system evolves. It seems that the developers are less in control of system evolution.

The latest MRs of Netbeans show an extreme concentration of bugs in a small percentage of CUs, as low as 8% of CUs holding 80% of bugs. We hypothesize that it is probably due to the fact that most CUs are not further changed during system evolution. We have to check this possibility in future work.

## 5 THREATS TO VALIDITY AND FUTURE WORK

The number of projects observed is limited and the AMs applied may be influenced by different factors which are not controlled during our analysis.

Our work involves Eclipse and Netbeans. To compare two software is not simple, and to judge which one has the better quality, is a challenge. There are different factors responsible for software quality, and the quality is not perceived by users in the same way.

When we use AMs, we can reach different level of customer satisfaction. Each software in fact have its specific audience and offers different mix of features and quality. For this reason in order to compare systems quality, we compare software products with analogue objectives. To reduce variability factors, we have decided to analyze software projects with similar attributes. Netbeans and Eclipse have the same target audience, and they compete for satisfy the same client requests and quality expectations. Both are programming environment, open source, developed in Java and with a wide user community. There are not other projects, of such size, which have been developed for as many years. In conclusion our sample is small, but it is homogeneous. By the way our study will be extended in the future.

The link between AMs and system faultiness is another threat to validity of our research. We can't describe how many bugs are avoid by AMs employment, but this is not our scope. We studied two softwares developed with AMs, and we compare their evolution and fault tendency. In this first work, our goal is to interpret how software evolution is justifiable with AMs employment.

Our statistic analysis is limited to mean values of CK metrics and bug number. We know that CK metric distribution follow in general a power-law (Concas *et al.* 2007), and for this reason the average is just a rough measure of the metrics. Thought, we believe that the average of these metrics can give an idea of the average quality of the system.

## 6 CONCLUSIONS

We presented an approach to extract and match information from code repositories, such as CVS and SVN, and bug tracking systems such as Bugzilla. The data about Java code were analyzed using a complex network approach, where classes are nodes and their relationships are edges. The software graph is also used to compute some CK metrics. Since bugs are found and fixed on software code files (Compilation Units), we introduced the concept of CU graph, and adapted CK metrics, originally introduced for classes, to cope with CUs.

We used the above approach to analyse two large open source Java projects in the same domain, Eclipse and Netbeans, to assess relationship among OO metrics and bug proneness. Overall, we found that both systems have a resolution process, that allows to resolve most issues when a main release is released, and that in both systems a small percentage of CUs holds most of issues, following the Pareto principle. This is an important result from the software engineering point of view. In fact, a review of a small fraction of faulty CU may have an exponential impact on the overall amount of software defects detectable and fixable.

We also found that Eclipse, the system developed following agile practices in a consistent way, has a foreseeable evolution of its metrics. On the other hand, Netbeans behavior is less foreseeable with respect to OO metrics and average number of issues per CU.

Future work will address in deeper detail the analysis of the relationship between CUs metrics and bug evolution, discriminating between CUs that were changed/added in the release and CUs unchanged. Moreover, the dynamic propagation of bugs along the software graph will be studied.

# REFERENCES

Abreu, F. B., 1995. The MOOD Metrics Set. In: Proc. ECOOP'95 Workshop on Metrics, Åarhus, Denmark.

Andersson C., Runeson P., 2007. A Replicated Quantitative Analysis of Fault Distributions in Complex Software Systems, IEEE Trans.Software Eng., vol. 33, no. 5, pp. 273-286.

Chidamber, S and Kemerer, C., 1998. A Metrics Suite for Object-Oriented Design. IEEE Trans. Software Eng., vol. 20, no. 6, pp. 476-493.

Concas, G., Marchesi M., Pinna S., Serra N., 2007. Power-Laws in a Large Object-Oriented Software System . IEEE Trans. Software Eng., vol. 33 no. 10, pp. 687-708.

Eclipse web site, 2009. http://www.eclipse.org.

Fischer M., Pinzger M., Gall H., 2003. Analyzing and relating bug report data for feature tracking. In Proc. 10th Working Conference on Reverse Engineering, WCRE'03, Victoria, British Columbia, Canada.

Junit web site, 2009. http://www.junit.org

Kim, S., Pan, K., Whitehead Jr., E.J., 2006. Micro Pattern Evolution. Proc. International Workshop on Mining Software Repositories (MSR'06), Shanghai, China.

Netbeans web site, 2009. http://www.netbeans.org

Purushothaman R., Perry D.E., 2005. Toward Understanding the Rhetoric of Small Source Code Changes, IEEE Trans.Software Eng., vol. 31, no. 6.

Śliwerski J., Zimmermann T., Zeller A., 2005. When do changes induce fixes?. Proc. International Workshop on Mining Software Repositories (MSR'05), St. Louis, Missouri, U.S.

Zhang H., 2008. On the Distribution of Software Faults, IEEE Trans.Software Eng., vol. 34, no. 2, pp. 301-302.

Zimmermann T., Nagappan N., 2008. Predicting Defects using Network Analysis on Dependency Graphs, In ICSE'08, Leipzig, Germany.